\documentclass[prb,twocolumn,showpacs,preprintnumbers,amsmath,amssymb]{revtex4}

\usepackage{graphicx}
\usepackage{dcolumn}
\usepackage{bm}

\preprint{submitted in PRB}

\begin{document}

 \title{Slater-Pauling Behavior and Origin of the Half-Metallicity of the Full-Heusler Alloys}

\author{I. Galanakis}\email{I.Galanakis@fz-juelich.de}
\author{P. H. Dederichs}
\affiliation{Institut f\"ur Festk\"orperforschung, Forschungszentrum J\"ulich,
D-52425 J\"ulich, Germany}
\author{N. Papanikolaou}
\affiliation{Fachbereich Physik, Martin-Luther Universit\"at, Halle-Wittenberg,
D-06099 Halle, Germany\\
Institute of Materials Science, NCSR ``Demokritos,'' 153~10 Aghia Paraskevi, 
Athens, Greece}

\date{\today}

\begin{abstract}
Using the full-potential screened  Korringa-Kohn-Rostoker method we study the
full-Heusler alloys based on Co, Fe, Rh and Ru. 
We show that many of these compounds show a half-metallic behavior, however in contrast to
the half-Heusler alloys the energy gap in the minority band is extremely small due to states 
localized only at the Co (Fe, Rh or Ru) sites
which  are not present in the half-Heusler compounds. 
The full-Heusler  alloys
show a Slater-Pauling behavior and the total spin-magnetic moment
per unit cell ($M_t$) scales with the total number of valence
electrons ($Z_t$) following the rule: $M_t=Z_t-24$. We explain why
the spin-down band contains exactly 12 electrons using arguments
based on the group theory and show that this rule holds  also 
for compounds with less than 24 valence electrons. Finally we discuss the deviations 
from this rule and the differences compared to the half-Heusler
alloys.
\end{abstract}

\maketitle

\section{Introduction \label{sect1}}
The increased interest in the field of magnetoelectronics or
spinelectronics during the last decade\cite{Wolf} intensified the
research on the so-called half-ferromagnetic materials. The latter
ones present a gap in the minority band and thus can be used as
perfect spin-filters or to enhance the performance of
spin-dependent devices as electrons at the Fermi level are 100\%
spin polarized. The first material which has been predicted to be
a half-ferromagnet was the half-Heusler alloy NiMnSb found by de Groot
and collaborators\cite{groot} in 1983. This prediction has been
verified also by other authors\cite{calcNiMnSb,iosif,iosif1} and the
half-ferromagnetic character has been also well established
experimentally both by using positron annihilation
experiments\cite{Hanssen} or inverse
photoemission.\cite{kirillova} Recently there is an increased
interest on thin films of this material both
experimentally\cite{films} and using first-principle
calculations.\cite{theoryNiMnSbfilms,iosif2}

\begin{figure}
 \includegraphics[scale=0.35]{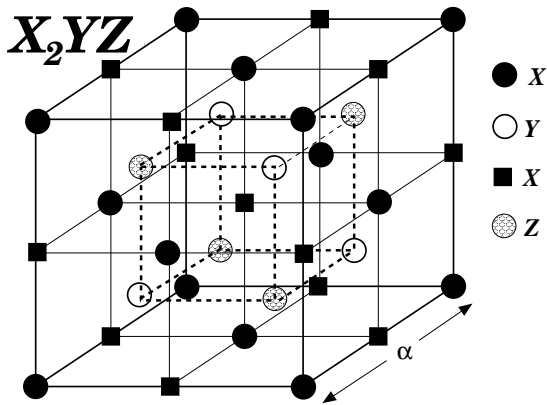}
  \caption{Schematic representation of the $L2_1$ structure. The lattice
consists of 4 fcc sublattices. The unit cell is that of a fcc
lattice with four atoms as basis: X at $(0\:0\:0)$ and
$({1\over2}\:{1\over2}\:{1\over2})$, Y at
$({1\over4}\:{1\over4}\:{1\over4})$  and Z at
$({3\over4}\:{3\over4}\:{3\over4})$  in Wyckoff coordinates.
\label{fig1}}
\end{figure}

Although the half-Heusler alloys like NiMnSb have attracted a lot
of interest, the second family of Heusler compounds, the so-called
full-Heusler alloys have been studied much more extensively due to
the existence of diverse magnetic phenomena,\cite{landolt,landolt2}
 mainly the transition from a ferromagnetic phase to an
antiferromagnetic one by changing the concentration of the
carriers.\cite{Kubler83} The full-Heusler alloys have the type
X$_2$YZ (see Fig.\ref{fig1}) and they crystallize in the $L2_1$
structure which consists of four fcc sublattices. Webster and
Ziebeck\cite{Webster} were the first to synthesize full-Heusler
alloys containing Co, and Ishida and
collaborators\cite{Ishida95,Fujii90} have proposed that the
compounds of the type Co$_2$MnZ, where Z stands for Si and Ge, are
also half-ferromagnets. 
Also the Heusler alloys of the type Fe$_2$MnZ have been proposed to
show half-ferromagnetism.\cite{Fujii95} But Brown \textit{et
al.}\cite{Brown00} using polarized neutron diffraction
measurements have shown that there is a finite very small
spin-down density of states (DOS) at the Fermi level instead of an 
absolute gap in agreement 
with the \textit{ab-initio} calculations of K\"ubler \textit{et al.} 
for the Co$_2$MnAl and Co$_2$MnSn compounds.\cite{Kubler83}
Recently, Ambrose \textit{et al.}\cite{Ambrose} managed to grow a Co$_2$MnGe
thin film on a GaAs(001) substrate by molecular beam
epitaxy and have proven the creation of domains during the growth.\cite{Yang}
Raphael \textit{et al.} have grown both thin films and 
single crystals of Co$_2$MnSi.\cite{Raphael}
Although these films were found to adopt the $L1_1$ structure there was 
a strong disorder between the Mn and Co sites even in the case of bulk Co$_2$MnSi.\cite{Ravel} 
Also, Geiersbach and collaborators
have grown (110) thin films of Co$_2$MnSi, Co$_2$MnGe and
Co$_2$MnSn using a metallic seed on top of a MgO(001)
substrate,\cite{Geiersbach}
Finally there also exist first-principles
calculations for the (001) surface of such an
alloy.\cite{iosif2,Ishida98} 

Suits\cite{Suits76} was the first to synthesize compounds of the
form Rh$_2$MnZ, where Z stands for Al, Ga, In, Tl, Ge, Sn and Pb.
They all crystallize in the $L2_1$ structure but the compounds
containing a II type $sp$ element show considerable disorder
between the $sp$ atom and the Mn site. They are all
ferromagnets and the compounds containing Ge, Sn and Pb
have a Curie temperature above room temperature.  Kanomata
\textit{et al.}\cite{Kanomata93} have grown crystals of
the type Ru$_2$MnZ, where Z stands for Si, Ge, Sn and Sb. Gotoh
\textit{et al.}\cite{Gotoh} have shown that these alloys are
antiferromagnets with N\'eel temperatures near room
temperature, and Ishida \textit{et al.}\cite{Ishida95-2} using
first-principles calculations demonstrated that the ground state is
antiferromagnetic with the Mn atoms in the (111) plane
being  antiferromagnetically coupled to the neighboring (111) planes.

In this contribution we study the full-Heusler alloys based on
Co, Fe, Ru and Rh by extending our work on the half-Heusler alloys
(see Ref.~\onlinecite{iosif1}) and on the transition metal monoarsenides
(see Ref.~\onlinecite{iosif3}). To perform the calculations we
have used the full-potential screened Korringa-Kohn-Rostoker (FSKKR)
Green's function method\cite{Pap02} in conjunction with the local
spin density approximation.\cite{Vosko} The details of our
calculations have been already described in
Ref.~\onlinecite{iosif1}. For all the compounds under study we
have used the experimental lattice
constants,\cite{landolt,Kanomata93} and have assumed that they are
all ferromagnets. In Section~\ref{sect2} we present the properties
of the Co$_2$MnZ compounds. In Section~\ref{sect3} we discuss the
origin of the gap in these compounds and the Slater-Pauling (SP) behavior 
of the total moments. In Section~\ref{sect4}
we present our results for some other interesting systems.
Finally in Section~\ref{sect5} we summarize our results and
conclude.

\section{C\lowercase{o}$_2$M\lowercase{n}Z compounds \label{sect2}}

\begin{table}
\caption{Calculated spin magnetic moments in $\mu_B$ using the experimental
lattice constants (see Ref.~\protect{\onlinecite{landolt}}) for the
Co$_2$MnZ compounds, where Z stands for the $sp$ atom.}
\label{table1} \begin{ruledtabular}
\begin{tabular}{rrrrrrr}
$m^{spin}$($\mu_B$) &  Co   &    Mn   & Z  & Total\\ \hline
Co$_2$MnAl    &  0.768  & 2.530 & -0.096 & 3.970  \\
Co$_2$MnGa    &  0.688  & 2.775 & -0.093 & 4.058 \\
Co$_2$MnSi    &  1.021  & 2.971 & -0.074 & 4.940 \\
Co$_2$MnGe    &  0.981  & 3.040 & -0.061 & 4.941 \\
Co$_2$MnSn    &  0.929  & 3.203 & -0.078 & 4.984
\end{tabular}
 \end{ruledtabular}
\end{table}

\begin{figure}
\includegraphics[scale=0.5]{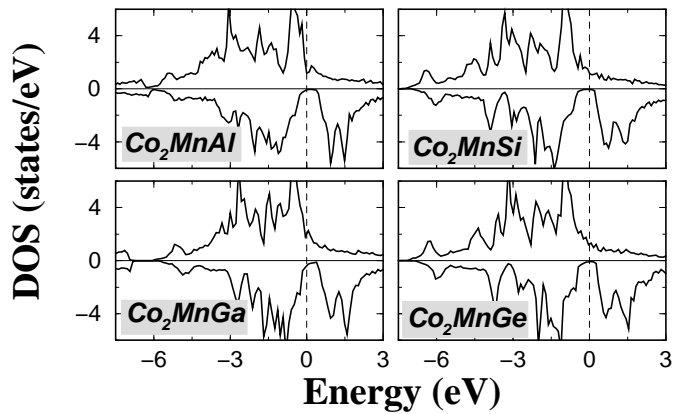}
 \caption{Calculated spin-projected DOS for the Co$_2$MnZ compounds, where
Z stands for Al, Ga, Si and Ge. They all posses a finite very small spin-down DOS
around the Fermi level. \label{fig2}}
\end{figure}

The first family of alloys, we will be interested in, are the compounds containing Co and Mn as they 
are the ones that have attracted most of the attention. They are all strong ferromagnets
with high Curie temperatures (above  600 K) and except the Co$_2$MnAl 
they show very little disorder.\cite{landolt}
They adopt the $L2_1$ structure, which we present in Fig.~\ref{fig1}. Each Mn or $sp$ 
atom has eight Co atoms as first neighbors sitting in an octahedral symmetry 
position, while each Co has four Mn and four $sp$ atoms as first neighbors and thus the 
symmetry of the crystal is reduced to the tetrahedral one.  The Co atoms occupying the 
two different sublattices are chemically equivalent as the environment of the one sublattice 
is the same as the environment of the second one but rotated by 90$^o$. 
The occupancy of two fcc sublattices by Co  (or in general by X) atoms distinguish the
full-Heusler alloys with the $L2_1$ structure from the half-Heusler compounds 
with the $C1_b$ structure, like \textit{e.g.} CoMnSb, where only one sublattice is occupied by Co atoms
and the other one is empty. Although in the $L2_1$ structure, the Co atoms are sitting 
on second neighbor positions, their interaction is important to explain the  
magnetic properties of these compounds as we will show in the next section. In 
Fig.~\ref{fig2} we have gathered 
the spin-resolved total density of states (DOS) for the Co$_2$MnAl, Co$_2$MnGa,
Co$_2$MnSi and Co$_2$MnGe compounds calculated using the FSKKR and in 
Table~\ref{table1} the
atom-projected and the total spin magnetic moment for these four compounds and 
for Co$_2$MnSn. Firstly as shown by photoemission experiments by Brown 
\textit{et al.}\cite{Brown98} in the case of Co$_2$MnSn and verified by our calculations
the valence band extends 5 eV below the Fermi level and the spin-up DOS shows a large peak
just below the  Fermi level for these compounds. Although Ishida 
\textit{et al.}\cite{Ishida95} have predicted them
to be half-ferromagnets with small spin-down gaps ranging from 0.1 to 0.3 eV 
depending on the material, within our calculations we find that the 
Fermi level falls within a  region of very small spin-down DOS for all these compounds.
Our results agree with the calculations of K\"ubler \textit{et al.}\cite{Kubler83} 
who  studied the Co$_2$MnAl and Co$_2$MnSn compounds using the Augmented Spherical Wave 
(ASW) method  and found also a very small spin-down DOS at the Fermi level and not 
a real gap. The reason of this pseudogap can be found in Fig.~\ref{fig3} 
where we have drawn the band structure
for the minority electrons in the case of the Co$_2$MnGe compound (our spin-down band structure is similar the
one obtained for Co$_2$FeGa and Mn$_2$VAl in Refs.~\onlinecite{deb} and \onlinecite{Weht} respectively). 
We see that the Fermi level
touches the highest occupied bands at the $\Gamma$ point and the lowest unoccupied bands
at the  X point and thus the indirect gap found in the half-Heusler 
alloys\cite{groot} is practically destroyed in these materials but there is  still a reasonably large direct gap
at the W, K and X points.  However we should mention that if we considerably enlarge the figure
with the band structure, it can be seen that the bands do not really touch the Fermi level
but there is a very small indirect gap of the order of 0.001 eV and thus the minimum of the 
minority unoccupied bands at   X and the maximum of the occupied bands at the $\Gamma$ point
are not degenerated. Our calculations include relativistic effects only within
the scalar-relativistic approximation, thus effects like the spin-orbit coupling can
lift the bands degeneracy and might even destroy the indirect gap. However we should mention
that in case of sufficient large band gaps like in the case of NiMnSb, the spin-orbit 
coupling does not destroy the half-metallicity.\cite{iosif1}

In the case of the half-Heusler alloys\cite{iosif1} like NiMnSb
the Mn spin magnetic moment is  very localized due to the exclusion of the spin-down 
electrons at the Mn site and amounts to about 3.7 $\mu_B$ in the case of NiMnSb. In the case of
CoMnSb the increased hybridization between the Co and Mn spin-down electrons decreased 
the Mn spin moment to about 3.2 $\mu_B$. In the case of the full-Heusler alloys each 
Mn atom has eight Co atoms as first neighbors instead of four as in CoMnSb and the
above hybridization is very important decreasing even further the Mn spin moment to less 
than 3 $\mu_B$ except in the case of Co$_2$MnSn where it is comparable to 
the CoMnSb compound.  The Co atoms are ferromagnetically coupled to the Mn
spin moments and they posses a spin moment that varies from $\sim$0.7 to 1.0 $\mu_B$, 
while the $sp$ atom has a very small negative moment which is one order of 
magnitude smaller than the Co moment. The negative sign of the induced $sp$ moment 
characterizes most of the studied full and half Heusler alloys with very few exceptions.

Another important point is that in half-metallic materials like the ones studied here the total 
spin moment should be an integer number since both the total number of valence electrons
as well as the number of occupied minority states are integers.
However our results in Table~\ref{table1} do not give integer  numbers for the total moments,
but slight deviations of about 0.05 $\mu_B$. This does not arise from incorrect space 
integration, as it \textit{e.g.} can occur in the atomic sphere approximation. 
In our implementation of the full-potential, the space is divided into Voronoi 
polyhedra,\cite{Pap02} which exactly fill up the space without any overlap, so that
the space integration is performed exactly. Rather the small deviations arise 
from an inherent feature of the KKR-Green's function method, which due to the 
$\ell_{max}$-cutoff violates the state normalization. A proper state 
counting leading to integer numbers for the total charges could only be achieved 
if all angular momenta up to $\ell_{max}=\infty$ would be 
included in the calculation, which is practically impossible in realistic cases.
The problem can be overcome by the application of the   Lloyd's formula,\cite{Lloyd} 
which contains an implicit summation over all angular momenta, thus yielding the
correct total charge and moment. Since the evaluation of  Lloyd's formula is a 
complex numerical problem, this is usually avoided arising to the above small inconsistencies.

Recently, a member of our group\cite{Rudi} has succeeded in implementing the 
 Lloyd's formula in our Green's function code and we have tested the case of 
 Co$_2$MnGe. The calculations give indeed  an integer total moment of  5 $\mu_B$
(instead of  4.941 $\mu_B$ in Table~\ref{table1}), and the (non-integer) local moments 
are slightly increased. Most of the charge adjustment occurs in the metallic majority band 
and the Fermi level is practically unchanged, situated as in Fig.~\ref{fig2} in the 
minority gap. This is also plausible from energetic point of view; the total energy 
favors this position of the Fermi level. Based on this experience and on calculations as 
above with different $\ell_{max}$ cut-offs, we conclude  (i) that in our calculations the correct
criterium for half-metallicity is, that the Fermi level is in the minority gap and (ii) that the
small deviations of the total moments from integer values are insignificant.

Thus we have verified by the DOS that all compounds under study in this section are 
half-metals.
The compounds containing Al and Ga have 28 valence electrons 
and the ones containing Si, Ge and Sn 29 valence electrons. The first compounds have 
a total spin moment of 4$\mu_B$ and the second ones of 5 $\mu_B$ which agree 
with the experimental deduced moments of these compounds.\cite{Dunlap} So it seems that 
the total spin moment, $M_t$, is given with respect to the total number of valence 
electrons, $Z_t$, from the simple relation: $M_t=Z_t-24$. In the following we will 
analyze the origin of this rule.

\section{Origin of the Gap and Slater-Pauling Behavior \label{sect3}}

\begin{figure}
\includegraphics[scale=0.45]{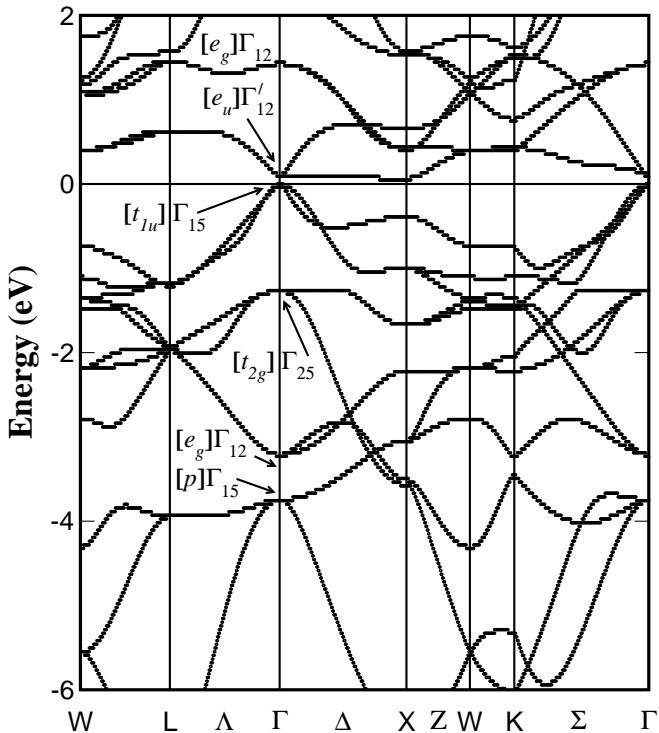}
  \caption{Spin-down band structure of the Co$_2$MnGe compound. The indirect
gap, present in the half-Heusler alloys, is practically destroyed. For the explanation
of the different representations of the group symmetry of the $\Gamma$ point look
at Table~\ref{table2}. In brackets we present the type of orbitals transforming following each representation 
(see Fig.~\ref{fig4}). \label{fig3}}
\end{figure}

As we mentioned above, the total spin magnetic moments of the 
Co$_2$MnZ compounds follow the $M_t=Z_t-24$ rule. A similar relation,
\textit{i.e.} $M_t=Z_t-18$, is also found for the half-Heusler 
compounds.\cite{jung,iosif1} Both state nothing more than the well known 
Slater-Pauling behavior.\cite{Kubler84} 
In such a picture the occupancy of the spin-down bands does not change and the extra or missing
electrons are taken care of by the spin-up states only.
The 24 means that there are 12 occupied spin-down states, as the total 
moment, which is the number of uncompensated spins, is given by the total
number of valence electrons $Z_t$ minus two times the number of minority electrons.

In Fig.~\ref{fig3} we present the representations of each one of the bands 
at the $\Gamma$ point (see Table~\ref{table2} for the different representations). 
Firstly the $sp$ atom creates one $s$ band and three $p$ 
bands which are fully occupied. The $s$ electrons transform following
the $\Gamma_1$ representation; we do not show this band in Fig.~\ref{fig3} 
as it very low in energy and it is well separated by the other bands. The $p$ electrons 
of the $sp$ atom transform following the $\Gamma_{15}$  representation and they 
hybridize with $p$ electrons of the Mn and  Co atoms which transform with the
same representation. As can be seen in the band structure, these bands are lower than 
the bands that have mainly $d$ character but they are not well separated by them
(there is a band crossing along the $\Gamma$K direction).
As in the half-Heusler alloys,\cite{iosif1} the 4 $sp$ bands can be only partially filled
by the $n$ valence electrons of the $sp$ atom ($n=3$ for Al, Ga or 4 for Si Ge and Sn),
so that an additional $8-n$ $d$ electrons are accommodated in these bands (4 $d$-electrons
in the case of Co$_2$MnGe or 5 $d$-electrons for Co$_2$MnAl). Therefore
in the Heusler alloys the effective number of $d$ electrons (in the higher lying $d$ 
bands) can be controlled by the valence of the $sp$ atom. This is a very unusual behavior 
for metallic systems, which can be used to engineer Heusler alloys with very different magnetic 
properties (see Section~\ref{sect4}).

\begin{table}
\caption{Representations  of the real space octahedral ($O_h$)
symmetry group (first column). In the second column the corresponding 
representations of the symmetry
group of the $\Gamma$ point following the  nomenclature introduced in Ref.
\protect{\onlinecite{Wigner}}.
In the third and fourth columns we present the orbitals which
transform following each one of the different representations.
Notice that the whole crystal has the tetrahedral $T_d$ symmetry but the
lattice consisted only of Co atoms has the $O_h$ symmetry;  $T_d$  is a subgroup of 
 $O_h$. Thus it is 
possible to have states located only at the Co sites, \textit{e.g.} 
the $d$ orbitals transforming according to the $E_u$ representation.  
Also the  $d$ hybrids transforming according to the $T_{1u}$ representation
are localized at the Co atoms as there are no $d$ states at the Mn site with the
same representation.  
The subscripts $\mathrm{a}$ and $\mathrm{b}$ refer to orbitals
at the two different Co sites in the unit cell (look Fig.~\protect{\ref{fig1}});
the 1, 2, 3, 4 and  5 refer to $d$ orbitals of the
$xy,\:yz,\:zx,\:3z^2-r^2$ and $x^2-y^2$ symmetries, respectively;
the  1, 2 and 3 refer to $p$ orbitals of the $x$, $y$ and $z$
symmetries, respectively.}
\label{table2}\begin{ruledtabular}
\begin{tabular}{rrcc}
 $O_h$ &  Ref.~\protect{\onlinecite{Wigner}} & Co-Co &  Mn or Ge \\ \hline
$A_{1g}$ & $\Gamma_1$            & $s_{\mathrm{a}}+s_{\mathrm{b}}$ & $s$ \\
$A_{1u}$  & $\Gamma_1^\prime$     & $s_{\mathrm{a}}-s_{\mathrm{b}}$ & \\
$E_g$     &  $\Gamma_{12}$        & $d_{i\mathrm{a}}+d_{i\mathrm{b}}$  [$i$=4,5]
& $d_4\:d_5$ \\
$E_u$     &   $\Gamma_{12}^\prime$ & $d_{i\mathrm{a}}-d_{i\mathrm{b}}$  [$i$=4,5]&\\
$T_{2g}$  &   $\Gamma_{25}$       & $p_{i\mathrm{a}}-p_{i\mathrm{b}}$ \&
$d_{i\mathrm{a}}+d_{i\mathrm{b}}$ [$i$=1,2,3] & $d_1\:d_2\:d_3$ \\
$T_{1u}$  &$\Gamma_{15}$ &  $p_{i\mathrm{a}}+p_{i\mathrm{b}}$ \&
$d_{i\mathrm{a}}-d_{i\mathrm{b}}$ [$i$=1,2,3]  & $p_1\:p_2\:p_3$
\end{tabular}
\end{ruledtabular}
\end{table}

In the case of the half-Heusler alloys, like CoMnSb, there is only  one Co atom per unit cell
and its $d$ valence electrons are hybridizing with the Mn ones creating five bonding states below
the Fermi level and five antibonding ones above the Fermi level. In the full-Heusler alloys 
the existence of the second Co atom makes the physics of these systems more complex.
As we mentioned above the whole crystal has the tetrahedral symmetry ($T_d$). But if we 
neglect the Mn and the $sp$ sites, then the Co atoms themselves sit on a cubic lattice 
respecting the octahedral symmetry ($O_h$). So there could be states obeying the $O_h$
being localized exclusively at the Co sites; note here that 
the $T_d$ is  a subgroup of $O_h$. Thus we will take into account firstly the 
interactions between the two inequivalent Co 
sites and then there interaction with the Mn or the $sp$ atom, as was also the 
case for the Fe$_2$MnZ compounds.\cite{Fujii95}

\begin{figure}
\includegraphics[scale=0.30]{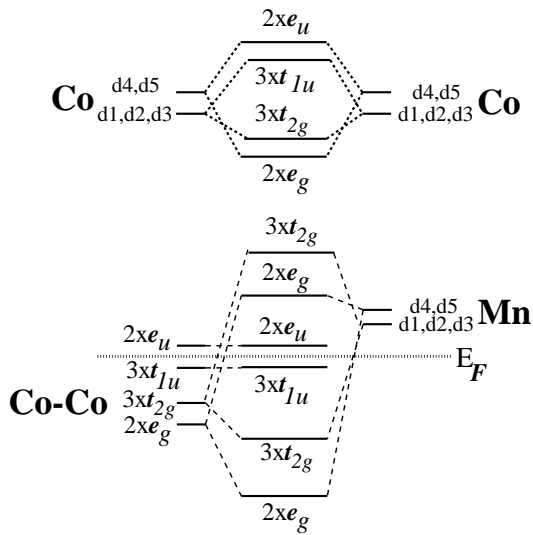}
  \caption{Possible hybridizations between spin-down orbitals sitting at different sites
in the case of the Co$_2$MnGe compound.
To explain the properties of the full Heusler alloys, firstly we consider the
hybridization between the two different Co atoms and afterwards the
hybridization with the Mn atom. The names of the orbitals follow the
nomenclature introduced in Table~\protect{\ref{table2}}. The coefficient represents
the degeneracy of each orbital. \label{fig4}}
\end{figure}

In order to discuss the behavior of the $d$ electrons in the full Heusler alloys
we have drawn schematically in Fig.~\ref{fig4} the possible hybridizations
between the different atoms. The $d_{1...5}$ orbitals correspond to the 
$d_{xy},\:d_{yz},\:d_{zx},\:d_{3z^2-r^2}$ and $d_{x^2-y^2}$ orbitals, respectively. 
The symbol $e_g$ means that the orbital transform following 
the $E_g$ representation.
Note that due to symmetry, the $e_g$ orbitals at the Co site can only couple with
 $e_g$ orbitals at the other Co site or at the Mn site. The same applies for the $t_{2g}$
orbitals.
Looking at Fig.~\ref{fig4} we see firstly that when two neighboring Co atoms interact,
their $d_4$ and $d_5$ orbitals form bonding $e_g$ and antibonding $e_u$ states; the
coefficient in front of each orbital is the degeneracy of this orbital. 
The $d_1,\:d_2$ and $d_3$ orbitals of each Co also hybridize creating
a triple degenerated bonding $t_{2g}$ orbital and a triple degenerated antibonding 
 $t_{1u}$ orbital.

\begin{figure}
\includegraphics[scale=0.5]{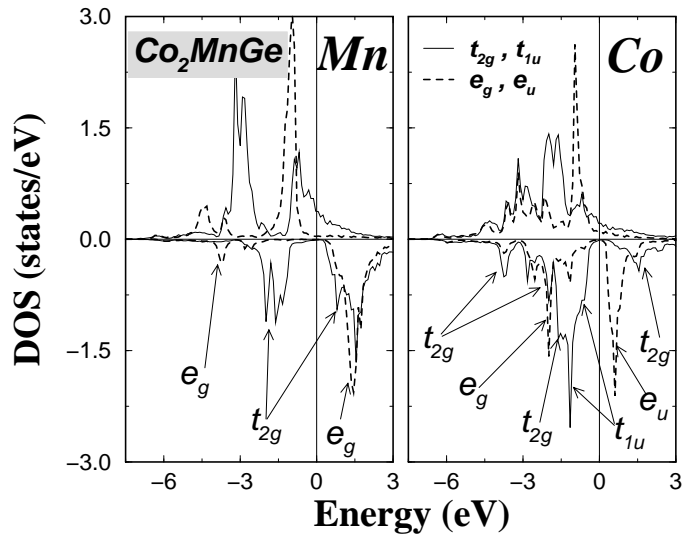}
  \caption{ Projected $d$ DOS on the double and on the triple degenerated representations for each atom
in the Co$_2$MnGe compound. We also
give the character of each peak for the spin-down states. Notice that in the minority bands around the Fermi level
there are only Co states.
\label{fig5}}
\end{figure}

As we show in the second part of Fig.~\ref{fig4},
the double degenerated $e_g$ orbitals hybridize with the $d_4$ and $d_5$ of the Mn
that transform also with the same representation. They 
create a double degenerated bonding $e_g$ state that is very low in energy and an antibonding 
one that is unoccupied and above the Fermi level. The $3\times t_{2g}$ Co orbitals
couple to the $d_{1,2,3}$  of the Mn and create 6 new orbitals, 3 of
which are bonding and are occupied and the other three are antibonding and 
high in energy. Finally the $2\times e_u$ and  $3\times t_{1u}$ Co orbitals cannot 
couple with any of the Mn $d$  orbitals as there are none transforming with the $u$   
representations. The $t_{1u}$ states are below the Fermi level and they are occupied while the
$e_u$ are just above the Fermi level. Thus in total 8 minority $d$ bands are filled and 7 are
empty. Our description is somewhat different from the 
one in Ref.~\onlinecite{Fujii95} where it has been assumed that 
the orbitals just below the 
Fermi level are also $t_{2g}$ and not $t_{1u}$ as in our case. To elucidate this 
difference  we have drawn in Fig.~\ref{fig5} the atomic-resolved $d$ DOS projected
on the double degenerated and the triple degenerated representations. Although we 
cannot distinguish in our projection the $t_{2g}$ from the $t_{1u}$ and the 
$e_g$ from the $e_u$, around the Fermi level the Mn atom present a broad spin-down
gap which is not present at the Co sites. So minority states around the 
gap are localized at the Co and do not couple to Mn, and the only states that 
have this property are the $t_{1u}$ and the $e_u$. Thus the peak below the Fermi level
is the  $3\times t_{1u}$ state and the peak just above the Fermi level is 
the   $2\times e_u$ state. This also explains why the gap is small. The two cobalt 
atoms are second neighbors and their hybridization is not so strong and the 
splitting of the states is small and thus the energy distance between the $t_{1u}$ levels and the
$e_u$ ones is small. As these states do not hybridize with the Mn states their splitting
does not change and the gap is considerably smaller than the one in the 
half-Heusler alloys. In the latter compounds we have only one Co atom per unit cell coupling to the Mn
atom and so the $t_{1u}$ and the $e_u$ states are absent and only the $e_g$ and 
$t_{2g}$ survive. Therefore a real gap exists in the half-Heusler alloys  and the minority 
valence and the minority valence bands contain 9 electrons: $1\times s,\: 3\times p$ and 
$5\times d$.

To summarize, in the case of the full-Heusler alloys we have 8 occupied minority $d$ states 
per unit cell: the double degenerated $e_g$ very low in energy, the triple degenerated
$t_{2g}$ orbital and finally the triple degenerated $t_{1u}$ just below the Fermi 
level. Thus in total we have 12 minority occupied states per unit cell, one with 
$s$ character, three with $p$ character and 8 with $d$ character. 
Therefore the total moment obeys the simple rule $M_t=Z_t-24$ as compared 
to $M_t=Z_t-18$ for the half-Heusler alloys.
Note here that as shown in Fig.~\ref{fig4} we have in total 15 spin-down $d$ states,
meaning 30 in total if we take into account both spin directions, so the states count
is correct as each of the two Co atoms and the Mn one contributes totally 10 $d$ states.
We can trace these states also in the spin-down band structure analyzing
the character of each band at the $\Gamma$ point. In Table~\ref{table2} we have included 
the representations of the symmetry group of the $\Gamma$ point in the reciprocal lattice 
using the nomenclature introduced in Ref.~\onlinecite{Wigner}.
The symmetry point group  of the $\Gamma$  has the same symmetry operations with the $O_h$.
Firstly as said above we have a $s$ like 
band not shown in the figure with a $\Gamma_1$ state at the $\Gamma$ point 
and then we find at $\Gamma$ a triple degenerated point that has the 
$\Gamma_{15}$ representation corresponding to the 
$p$ like orbitals. Above this point there  is a double-degenerated $\Gamma_{12}$ point
which corresponds to the $e_g$ orbitals while the other $e_g$ orbitals for Co$_2$MnGe
are found above the Fermi level and also above the unoccupied $e_u$ orbitals 
that correspond to the double degenerated point with the $\Gamma_{12}^\prime$ symmetry.
Finally, there are  two triple degenerated points $\Gamma_{25}$ and $\Gamma_{15}$ 
which  correspond to the occupied $t_{2g}$ and $t_{1u}$ orbitals, respectively, 
while the other unoccupied  $t_{2g}$ orbitals ($\Gamma_{25}$) are high in energy and are not shown
in the figure.

From the above discussion we find that in the minority band 7 $d$ states above $E_F$ are unoccupied.
Thus the largest possible moment, which a full-Heusler alloys can have, is 7 $\mu_B$, since
in this case all majority $d$ states are filled. This is different from the half-Heusler compounds 
which have five empty $d$-states in the minority band and therefore   a maximum moment of 5 $\mu_B$.

\begin{figure}
\includegraphics[scale=0.5]{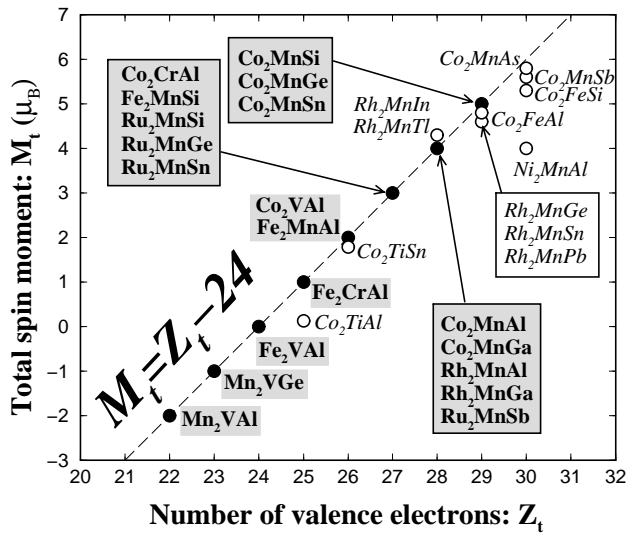}
\caption{Calculated total spin moments for all the studied Heusler alloys.
The dashed line represents the Slater-Pauling behavior. With open
circles we present the compounds deviating from the SP curve. To decide whether one alloy
is half-ferromagnet or not, we have used the DOS and not the total spin-moments 
(see Section~\ref{sect2}). \label{fig6}}
\end{figure} 
 
\section{Other Full-Heusler compounds following the SP curve
\label{sect4}}

Following the discussion of the previous section we will go on investigating 
other full-Heusler alloys that can follow  the Slater-Pauling curve and 
in Fig.~\ref{fig6} we have plotted  the total spin magnetic moments for all
the compounds under study as a function of the total number of valence electrons.
The dashed line  represents the rule: $M_t=Z_t-24$.  In the following we will 
analyze all these results. Overall we see that many of our results coincide with 
the Slater-Pauling curve. Some of the Rh compounds show small deviations which are
more serious for the Co$_2$TiAl compound. We see that there is no compound with a total
spin moment of 7 $\mu_B$ or even 6 $\mu_B$. Moreover we found  also examples of  
half-metallic materials with less than 24 
electrons, Mn$_2$VGe with 23 valence electrons and Mn$_2$VAl with 22 valence electrons. 

\subsection{Co$_2$YAl and Fe$_2$YAl compounds \label{subsect1}}

\begin{table}
\caption{Calculated spin magnetic moments in $\mu_B$ using the experimental
lattice constants (see Ref.~\protect{\onlinecite{landolt}}) for the
full-Heusler alloys containing Co, Fe and Mn.}
\label{table3}\begin{ruledtabular}
\begin{tabular}{rrrrrrr}
$m^{spin}$($\mu_B$) &  Co,Fe,Mn   &    Y   & Al,Si,Ge  & Total\\ \hline
Co$_2$TiAl    &  0.072  &-0.013 & -0.002 & 0.130 \\
Co$_2$TiSn      &  0.911  &-0.039 &  0.001 & 1.784 \\
Co$_2$VAl     &  0.863  & 0.232 & -0.033 & 1.926 \\
Co$_2$CrAl    &  0.755  & 1.536 & -0.091 & 2.955 \\
Co$_2$MnAl    &  0.768  & 2.530 & -0.096 & 3.970  \\
Co$_2$FeAl    &  1.129  & 2.730 & -0.099 & 4.890\\
Fe$_2$VAl     &  \multicolumn{4}{c}{paramagnet} \\
Fe$_2$CrAl    & -0.093 & 1.108 &-0.011 & 0.910 \\
Fe$_2$MnAl    & -0.275  & 2.548 & -0.019 & 1.979 \\
Fe$_2$MnSi    &  0.191   & 2.589 & -0.029 & 2.943  \\
Mn$_2$VAl     & -1.413  & 0.786& 0.018 & -2.021 \\
Mn$_2$VGe     &  -0.750 & 0.476 & 0.021 & -1.003
\end{tabular}
 \end{ruledtabular}
\end{table}

\begin{figure}
\includegraphics[scale=0.48]{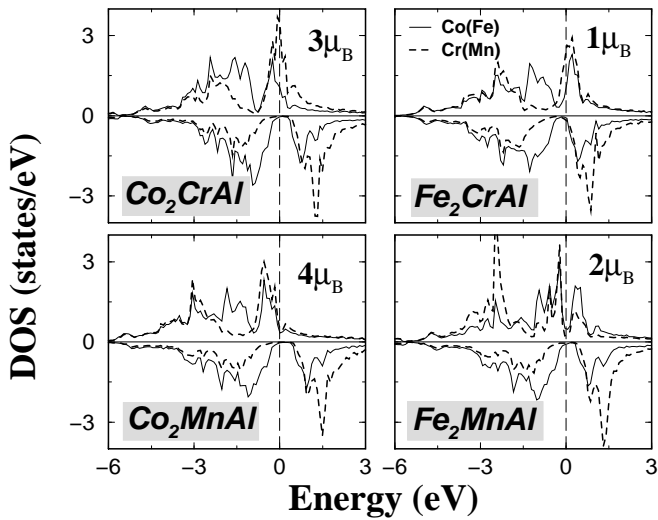}
\caption{Calculated atom- and spin-projected DOS for the Co(Fe)$_2$Mn(Cr)Al
compounds. They all present a spin-down pseudogap. The numbers give the total moments.\label{fig7}}
\end{figure}

We have calculated the spin moments of the compounds Co$_2$YAl where Y= Ti, V, Cr, Mn 
and Fe and in Table~\ref{table3} we have gathered the atomic and the total spin 
magnetic moments. There are  experimental results only for the moment at the Co site 
for the Ti, V, and Cr compounds using hyperfine field measurements by Pendl 
\textit{et al.}\cite{Pendl}  and  by Carbonari
\textit{et al.}\cite{Carbonari},  which agree very well with our 
\textit{ab-initio} results.
The compounds containing V, Cr and Mn show a similar behavior.
As we substitute Cr for Mn, that has one valence electron less than Mn, 
we depopulate one Mn spin-up state  and thus the spin moment of Cr is around 
1 $\mu_B$ smaller than the Mn one while the Co moments are practically the same for 
both compounds.
This behavior is clearly seen in Fig.~\ref{fig7} where we present the 
atom- and spin-resolved DOS for the two compounds. The minority DOS is the 
same for both compounds as they follow the SP curve and this is also the case 
for the Co spin-up DOS. In the case of the Cr compound the Fermi level 
falls within a broad and large Cr spin-up peak. When we substitute Mn for Cr this peak 
moves lower in energy to account for the extra electron and the Fermi level
is now at the right edge of the  peak, but nothing else changes in the calculated DOS.
Substituting V for Cr has a larger effect since  also the Co spin-up DOS changes 
slightly and the Co magnetic moment is increased by about 0.1 $\mu_B$ compared to the
other two compounds and V possesses a small moment of ~0.2 $\mu_B$. This change in
the behavior is due to the smaller hybridization between the Co atoms and the V 
compared to the Cr and Mn atoms. Although all three  Co$_2$VAl, Co$_2$CrAl and 
Co$_2$MnAl compounds are on the SP curve as can be seen in  Fig.~\ref{fig6}, this is not 
the case for the compounds containing Fe and Ti. If the substitution of Fe for Mn
followed the same logic as the one of Cr for Mn then the Fe moment should be around
3.5 $\mu_B$ which is a very large moment for the Fe site.  Therefore it is energetically 
more favorable for the system that also the Co moment is increased, as it was also
the case for the other systems with 29 electrons like Co$_2$MnSi, but while 
the latter one makes it to 5 $\mu_B$, Co$_2$FeAl reaches a value of 4.9 $\mu_B$.
A similar behavior was seen also in the case of the isoelectronic Co$_2$FeGa compound, 
but the total spin moment was slightly larger than 5  $\mu_B$.\cite{deb} 
In the case of Co$_2$TiAl, is is energetically more favorable to
have a weak ferromagnet than an integer moment of 1 $\mu_B$ as it is
very difficult to magnetize the Ti atom. Even in the 
case of the Co$_2$TiSn the calculated total spin magnetic moment of  1.78 $\mu_B$
(compared to the experimental value of 1.96  $\mu_B$)\cite{Engen} 
arises only from the Co atoms as was also shown experimentally by Pendl 
\textit{et al.},\cite{Pendl} and the Ti atom is practically paramagnetic and the 
latter compound fails to follow the SP curve. 

As a second family of materials we have calculated also the compounds containing Fe 
and we present their total spin moments  also in Table~\ref{table3}.
Fe$_2$VAl has in total 24 valence electrons and is a semimetal, \textit{i.e.} 
paramagnetic with a very small DOS at the Fermi level, as 
it is already known experimentally.\cite{Fe2VAl} 
All the studied Fe compounds follow the SP behavior as can be seen in Fig.~\ref{fig6}. 
In the case of the Fe$_2$CrAl and Fe$_2$MnAl compounds the Cr and Mn atoms have spin 
moments comparable to the Co compounds and similar DOS as can be seen in 
Fig.~\ref{fig7}. In order to follow the SP curve the Fe in  Fe$_2$CrAl is practically
paramagnetic while in Fe$_2$MnAl it has a small negative moment. 
When we substitute Si for Al in Fe$_2$MnAl, the extra electron exclusively populates Fe spin-up 
states and the spin moment of each Fe atom is increased by 0.5 $\mu_B$ 
contrary to the corresponding Co compounds where also the Mn spin moment was 
considerably increased. 

Finally we calculated as a test Mn$_2$VAl and Mn$_2$VGe which 
 have 22 and 23 valence electrons, respectively, 
to see if we can reproduce the SP behavior not only  for compounds with more than 24,
 but also for compounds with less than 24 electrons. As we have already shown Fe$_2$VAl is paramagnetic 
and Co$_2$VAl, which has two electrons more, has a spin moment of 2 $\mu_B$. 
Mn$_2$VAl has two valence electrons less than Fe$_2$VAl and as 
we show in Table~\ref{table3} its total spin moment is 
$-\:2\:\mu_B$, in agreement with previous \text{ab-initio} results,\cite{Weht}
 and thus it follows the SP behavior. To our knowledge there is no
compound with 23 valence electrons, which has been studied experimentally, so 
we decided to calculate Mn$_2$VGe using the lattice constant of Mn$_2$VAl. We have 
chosen this compound, because as can be seen in Ref.~\onlinecite{landolt} 
the compounds containing Al and Ge have practically the same lattice constants.
We found that adding one electron to  Mn$_2$VAl results in a decrease of
the absolute value of both the Mn and V spin moments (note that V and Mn are 
antiferromagnetically coupled) so that the resulting Mn$_2$VGe total spin magnetic 
moment is -1 $\mu_B$ following the SP curve as can be also seen in Fig.~\ref{fig6}. 

\subsection{The Ru and Rh compounds  \label{subsect2}}

\begin{table}
\caption{Calculated atom-resolved and total spin magnetic moments in $\mu_B$
using the experimental
lattice constants for the full-Heusler alloys containing Rh and Ru
(see Ref.~\protect{\onlinecite{landolt}} for the lattice constants of the Rh compounds
and Ref.~\protect{\onlinecite{Kanomata93}} for the Ru compounds).}
\label{table4}\begin{ruledtabular}
\begin{tabular}{rrrrrrr}
$m^{spin}$($\mu_B$) &  Ru, Rh   &    Mn   & Z  & Total\\ \hline
Ru$_2$MnSi    &  0.028 & 2.868 &  0.025 & 2.948 \\
Ru$_2$MnGe    &  0.002 & 2.952 & 0.021  & 2.977 \\
Ru$_2$MnSn    &  -0.051 & 3.137 & -0.001   & 3.034\\
Ru$_2$MnSb    &  0.222 & 3.495 & 0.018 & 3.957 \\
Rh$_2$MnAl    &  0.328 & 3.388 & -0.041 & 4.004 \\
Rh$_2$MnGa    &  0.312  & 3.461 & -0.033 & 4.052 \\
Rh$_2$MnIn    &  0.269 & 3.720 & -0.034 & 4.223 \\
Rh$_2$MnTl    &  0.266 & 3.765 & -0.027 & 4.270 \\
Rh$_2$MnGe    &  0.421  & 3.672 &  0.011 & 4.525 \\
Rh$_2$MnSn    &  0.393 & 3.831 & -0.010 & 4.607 \\
Rh$_2$MnPb    &  0.383 & 3.888 & -0.009 & 4.644
\end{tabular}
\end{ruledtabular}
\end{table}

\begin{figure}
\includegraphics[scale=0.48]{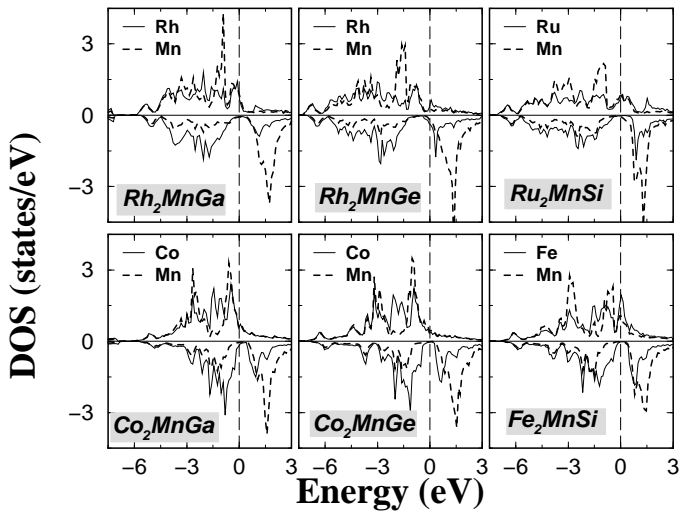}
\caption{Calculated atom- and spin-projected DOS for some of the Heusler alloys containing
Ru and Rh compared to the alloys containing Fe or Co that are isoelectronic to Ru and Rh,
respectively. In the case of the Rh or Ru compounds, the hybridization with the spin-down
Mn states is smaller resulting in larger Mn spin moments (see Table~\ref{table4}).
 \label{fig8}}
\end{figure}

To investigate further the Slater-Pauling behavior of the full-Heusler alloys we 
studied the ones containing a 4$d$ transition metal atom. 
As we have already mentioned in Section~\ref{sect1} the Ru compounds are 
antiferromagnets with N\'eel temperatures that reach room temperature.
We have calculated their properties assuming that they are ferromagnets and present
the calculated spin-magnetic moments in Table~\ref{table4}.  The Ru$_2$MnSi,
Ru$_2$MnGe and Ru$_2$MnSn have a total spin magnetic moment of 3 $\mu_B$ and 
Ru$_2$MnSb a moment of 4  $\mu_B$ following the rule for the magnetic moments
that we have already shown for the Co and Fe compounds and thus the Fermi level falls within the 
pseudogap contrary to the calculations in Ref.~\onlinecite{pugacheva} where the Fermi level
was above the gap. In the case of the alloys with Si, Ge
and Sn the Ru atom has  a practically  zero spin moment and  the total moment is 
carried by the Mn atoms. In Fig.~\ref{fig8} we have drawn the atomic and spin DOS 
for the Ru$_2$MnSi compound compared to the isoelectronic Fe$_2$MnSi compound.  
We see clearly from the DOS that the hybridization between the Mn and Ru 
spin-down states is smaller than in the case of the Fe compound resulting 
in a larger Mn spin moment. Although Ru has a practically zero spin moment, we see that
the Fermi level falls within a broad peak of spin-up DOS. In the case of Ru$_2$MnSb
this peak is completely occupied resulting in an important induced spin moment at the
Ru site that couples ferromagnetically to the Mn one. 

The next family of compounds that we have studied are the 
ones containing Rh and Mn.\cite{Suits76,Jha85} 
In Table~\ref{table4} we present the calculated
spin magnetic moments. As can be seen in Fig.~\ref{fig8} the hybridization
between the Rh and the Mn spin-down states is smaller than in the case of
the isoelectronic Co compounds, \textit{i.e.} there are Mn states in the Co compound
that become Rh states in the Rh compound, thus leading to an increase of the 
Mn moment, and a  decrease of the Rh moment compared to the 
Co spin moment. This phenomenon is quite intense as the Mn moment increases 
in all cases  more than 0.6 $\mu_B$. From the studied compounds only Rh$_2$MnAl 
and Rh$_2$MnGa are 
exactly on the SP curve presented in Fig.~\ref{fig6}. The Rh$_2$MnIn and Rh$_2$MnTl
that are isoelectronic to the two previous compounds have a total spin moment of 
around 4.2 - 4.3 $\mu_B$, thus the Fermi level is slightly below the pseudogap in these
compounds. In the case of Rh$_2$MnGe, Rh$_2$MnSn and  Rh$_2$MnPb, that posses 29 valence
electrons, the total spin moment is around 4.6 $\mu_B$ slightly smaller than the ideal
5 $\mu_B$ and the Fermi level is slightly above the pseudogap. This is probably
due to the considerably larger lattice constant of the Rh compounds with respect to the 
isoelectronic Co ones. But in general, as can be seen also in Fig.~\ref{fig6}, where 
we summarize all our results, all the compounds are not very far from the SP curve 
and the deviations are small.

\subsection{Compounds with 30 valence electrons  \label{subsect3}}

As stated in Section~\ref{sect3} the maximal moment of a full-Heusler alloy
is 7 $\mu_B$, and should occur, when all 15 majority $d$ states are occupied. 
Analogously for a half-Heusler alloy the maximal moment is 5 $\mu_B$. However this limit
is difficult to achieve, since due to the hybridization of the $d$ states with
empty  $sp$-states of the transition metal atoms (sites X and Y in Fig.~\ref{fig1}), $
d$-intensity is transferred into states high above $E_F$, which are very difficult to
occupy. While we could identify in a recent paper on half-Heusler alloys (Ref.~\onlinecite{iosif1})
systems with a moment of nearly 5 $\mu_B$, the hybridization is much stronger
in the full-Heusler alloys so that a total moment of 7 $\mu_B$ seems to be impossible. Therefore
we restrict our search to possible systems with 6 $\mu_B$, \textit{i.e.} systems with 30 valence electrons.
We have studied some of the possible candidates and we present 
our results in Table~\ref{table5}. One obvious way to reach the 30 electrons
is to substitute, \textit{e.g.} in Co$_2$MnAl, Co by Ni, but Ni is practically paramagnetic and cannot carry
a large spin moment and thus the total spin magnetic moment of Ni$_2$MnAl is only 
4 $\mu_B$ far away from the ideal 6 $\mu_B$. The second way to achieve 30 electrons is
to use Fe at the Y site as it the case for the Co$_2$FeSi compound. 
Already Co$_2$FeAl was not reaching the 5 $\mu_B$ and adding one more electron 
can not  increase the total spin moment by more than 1 $\mu_B$. Although the Co moment reaches the 1.3 $\mu_B$
the Fe moment stays unchanged and the total spin moment is increased only by $\sim$0.4
 $\mu_B$ reaching the 5.3 $\mu_B$ instead of the ideal 6 $\mu_B$. 

\begin{table}
\caption{Calculated spin moments for full-Heusler alloys containing 30 valence
electrons per unit cell. The experimental lattice parameters
were taken from Ref.~\protect{\onlinecite{landolt}}.}
\label{table5}\begin{ruledtabular}
\begin{tabular}{rrrrrrr}
$m^{spin}$($\mu_B$) &  X   &    Y   & Z  & Total\\ \hline
Ni$_2$MnAl    &  0.364  & 3.359 & -0.062 & 3.973 \\
Co$_2$FeSi    &  1.271  & 2.756 & -0.031 & 5.268 \\
Co$_2$MnSb    &  1.113  & 3.401 & -0.007 & 5.620\\
Co$_2$MnAs    &  1.219  & 3.309 &  0.035 & 5.782     
\end{tabular}
\end{ruledtabular}
\end{table}

Our last test cases are the  Co$_2$MnSb and Co$_2$MnAs compounds. We have calculated 
 Co$_2$MnSb using the lattice constant of Co$_{1.5}$MnSb
as Co$_2$MnSb does not really exist. Adding Co to  Co$_{1.5}$MnSb results to the
creation of a Co rich phase.  Co$_2$MnSn has a total spin moment of 5  $\mu_B$. The 
additional electron increases both the Co and Mn spin moments and the total moment is
now 5.6  $\mu_B$. To our knowledge there is no experimental work on Co$_2$MnAs 
and we have calculated it
using the lattice constant of Co$_2$MnGe. This lattice constant should be very close
to the real one as also substituting Ga for Ge only marginally changes it. As shown in 
Table~\ref{table5} the calculated total spin moment is 5.8 $\mu_B$. But for 
both compounds if we increase their lattice constant by 4\% the Fermi level moves 
deeper in energy, as was the case also for the half-Heusler alloys,\cite{iosif1} and
now it falls within the pseudogap and the total spin moment for both of 
them reaches the ideal value of 6 $\mu_B$.  So
if both Co$_2$MnSb and Co$_2$MnAs can be grown on top of a substrate with the appropriate 
lattice constant using 
a technique like Molecular Beam Epitaxy, it is possible to get a material with a total
spin moment of 6  $\mu_B$ where the Fermi level will be within the pseudogap. 
In such a case, of course, there is the possibility that the lattice parameter along
the growth axis is contracted to account for the large in-plane lattice parameter, which
can lead to a change of the total spin moment.

\section{Conclusions\label{sect5}}

Using the full-potential screened  Korringa-Kohn-Rostoker method we studied the
full-Heusler alloys containing Co, Fe, Rh and Ru. We have shown using the 
scalar-relativistic approximation that for
all these compounds the top edge of the highest occupied spin-down
band and the bottom edge of the lowest unoccupied spin-down band
touch the Fermi level practically destroying the indirect gap. These compounds
show a Slater-Pauling behavior and the total spin-magnetic moment
per unit cell ($M_t$) scales with the total number of valence
electrons ($Z_t$) following the rule: $M_t=Z_t-24$. The Co-Co hybridization is 
primordial to  explain why the spin-down band contains exactly 12 electrons
and why only a tiny gap exists in these compounds.
Finally we have shown that it is possible to find the Slater-Pauling behavior 
even for materials with less than 24 valence electrons like Mn$_2$VAl and Mn$_2$VGe, 
 and that the compounds with 30 valence 
electrons are unlikely to achieve a total spin moment of 6 $\mu_B$.

\begin{acknowledgments}   
The authors acknowledge financial support from
the RT Network of {\em Computational Magnetoelectronics} (contract
RTN1-1999-00145) of the European Commission. We thank Dr. Rudi Zeller 
for providing us with a version of the KKR code incorporating the 
Lloyd's formula and for helpful discussions.
\end{acknowledgments}

\end{document}